# Urban form and COVID-19 cases and deaths in Greater London: an urban morphometric approach

**Alessandro Venerandi\***, **Luca Maria Aiello**[†], **Sergio Porta\***

\* Urban Design Studies Unit (UDSU), Department of Architecture, University of Strathclyde, Glasgow, UK. 75 Montrose St, Glasgow G1 1XJ, UK

[†] Department of Computer Science, IT University of Copenhagen, Copenhagen, DK
Rued Langgaards Vej 7, DK-2300 Copenhagen S, Denmark

Corresponding author's email: alessandro.venerandi@strath.ac.uk

## Abstract

The COVID-19 pandemic generated a considerable debate in relation to urban density. This is an old debate, originated in mid 19[th] century's England with the emergence of public health and urban planning disciplines. While popularly linked, evidence suggests that such relationship cannot be generally assumed. Furthermore, urban density has been investigated in a spatially coarse manner (predominantly at city level) and never contextualised with other descriptors of urban form. In this work, we explore COVID-19 and urban form in Greater London, relating a comprehensive set of morphometric descriptors (including *built-up density*) to COVID-19 deaths and cases, while controlling for socioeconomic, ethnicity, age, and co-morbidity. We describe urban form at individual building level and then aggregate information for official neighbourhoods, allowing for a detailed intra-urban representation. Results show that: (i.) control variables significantly explain more variance of both COVID-19 cases and deaths than the morphometric descriptors; (ii.) of what the latter can explain, *built-up density* is indeed the most associated, though *inversely*. The typical London neighbourhood with high levels of COVID-19 infections and deaths resembles a suburb, featuring a low-density urban fabric dotted by larger free-standing buildings and framed by a poorly inter-connected street network.

# Introduction

## The post-pandemic city: designing for the permanent wartime

Since December 2019, when the first cases of COVID-19 were reported in the Chinese city of Wuhan, a narrative of emergency has colonised the public discourse globally. Nearly all areas of science have contributed their insights into a phenomenon that has profoundly impacted our societies beyond the boundaries of medicine and public health. Urban planners and designers are no exception, as the pandemic continues to fuel a considerable wave of speculations about the "opportunities" associated with re-imagining and re-designing the post-pandemic city. The pandemic's impact on many aspects of urban life and functioning have been explored and reconsidered, from wellness and quality of life / urban compactness (Mouratidis, 2021, 2022) to mobility and travel behaviour (Wei et al., 2021), political and economic assets (Glaeser, 2021), the urban planning profession, resilience and disaster management (Allam and Jones, 2020), socio-economic factors (Qiu et al., 2020) and, of course, population density (Lin et al., 2020) among many others, covering a considerable range of scales from urban to metropolitan, regional and beyond, without excluding the historical perspective (Eltarabily and Elgheznawy, 2020).

During the first industrial revolution (Rifkin, 2011) a similar wave of intellectual and organisational thrill innovated cities seeking new social and political configurations in a world shaken by remarkable changes in class-relations, mode of production and distribution of wealth. Integral to the then innovative concept of "public health", urban planning emerged as one societal response to the "housing question" (Engels and Dutt, 1935), the horrifying living conditions of the working-class amassed in booming industrial cities and the disastrous infectious epidemics that soon followed. The *design policy* response ("Public Health Act" of 1848) came up sound and clear: density had to be drastically reduced. Since then, the "less density = healthier cities" equation was forever installed in the heart of the new profession of urban planning(Shulman, 2005). There it remains up until now, deeply influencing mainstream planning policies and practices globally, despite the aspirations of decades of place-making *"sustainable compact counter-revolution"* (Porta and Romice, 2014). Even though evidence is all but univocal (Hamidi et al., 2020), the "less density-more health equation" continues to capture planners.

In the aftermath of the 2003 SARS epidemic, the "Team Clean" commission nominated by the Hong Kong government investigated the connection between urban design and the spread of SARS in Hong Kong, without finding any (Clean Team, 2003). Regardless, they concluded that *"good urban design can contribute to a healthy lifestyle and environment"* (idem: 6), suggesting the creation of "breezeways" to improve air circulation by reducing density. Similarly, Ng et al. (Ng et al., 2005: 6) proposed road widening, more open spaces and re-aligning buildings according to main wind directions.

Even though a link with public health has always remained part of the urban planning remit, with the COVID-19 pandemic it is now back to centre-stage in the debate on the post-pandemic city, and the "density equation" with it. From January to March 2021, the Town Planning Review journal dedicated three consecutive special issues to urban planning and COVID-19, the first of which focuses on housing and urban form (Dockerill et al., 2021). As Webster nicely put it, urban planning is coming back to its core-constituent role of public-health density-focused intervention (2021: 123). Despite the complexity of the relation between city, community, and health (Duhl et al., 1999), the cheered return of urban planning back to its public health home (Chigbu and Onyebueke, 2021; D'Alessandro et al., 2017; Roe, 2019) does not come without risks of oversimplifications. In November 2020, Andres Duany, the father and founder of the Congress of New Urbanism (CNU), a think-tank supporting the compact European traditional city model, delivered a talk (Duany, 2020) against principles that pre-modern cities have been delivering for millennia and CNU preached for decades: density, face-to-face retail, and community engagement in decision-making.

## Are urban form and urban design relevant to COVID-19?

Density in cities is far less straightforward a concept than one may think (Angel et al., 2021). In the early 1960s, Jacobs (Jacobs, 1992) pointed out that *built-up density*, i.e., the number of buildings per hectare of developed land area, is not to be confused with *crowding*, i.e., the amount of people per habitable room (Moroni, 2016). It is on the ground of this confusion that urban planning policies of systematic slum clearance have been justified for decades, along with the displacement of communities living in conditions of extreme deprivation, including overcrowding. This malpractice, which still threatens contemporary slums in the Global South, appears reinvigorated under the new COVID-19 justification (Corburn et al., 2020; Jasiński, 2021). After evidence has been raised in decades of place-making research, that urban form may actively contribute to important urban dynamics including prosperity (Ewing and Hamidi, 2015; Porta et al., 2012; UN-Habitat, 2013; Venerandi et al., 2018) and quality of life (Romice et al., 2017; Venerandi et al., 2016), urban designers risk to inadvertently over-indulge in spatial determinism while tackling the COVID-19 emergency. Available scientific evidence relating urban form and communicable diseases, as it turns out, is not very abundant nor univocal. Sharifi and Khavarian-Garmsir (2020) found that limited evidence on this matter before COVID-19. After the COVID-19 outbreak, most relevant literature (61%) focuses on urban ecology, socio-economic inequality, and urban resilience. Aspects of urban management, governance and transport are also represented (28%), while urban form/design accounts for 5% only (idem: 3). The resulting evidence is found *"contrasting and inconclusive"* (idem: 10)*,* as sociodemographic, inequality and cultural-ethnic characteristics may stand in the way of the simple density equation. One of the most comprehensive of such studies (Hamidi et al., 2020) focuses on 900+ metropolitan counties in the USA and finds that, while the population *size* of metro areas is directly correlated with COVID-19 cases and deaths, *population density* is not. Fang and Wahba (2020) focus on 284 Chinese cities and report no significant correlation between population density and COVID-19 infection rate as of April 2020, measured at city level. They find, however, that denser cities are also wealthier, which may explain why they tend to show lower infections rates as they can support better public health strategies and infrastructures. In a study on 1,759 urban USA counties, covering 93% of overall national population, Carozzi (2020) find that denser areas tend to be hit harder earlier after the outbreak, but do react better in the longer term, resulting in no links between COVID-19 death rates and population density when time is considered.

In conclusion, available literature does not support the claim that "density" in cities relates, as such, to the spread or severity of communicable diseases, including COVID-19. However, it also shows gaps: 1) it is not based on a detailed and systematic description of urban form: rather, it overwhelmingly focuses on *population* density only (which can hardly be defined an urban form metric in the first place); 2) it nearly exclusively matches large-scale of extents (regional and national) with large-scale of unit of information (municipal, county or higher); 3) thus, it does not address the diversity of contemporary urban form at the tissue level (Kropf, 2018). This paper aims at filling these gaps by studying the relationship between urban form and COVID-19 tested-positive cases and deaths in Greater London, where urban form is kept centre-stage. In the description of urban form, this study uniquely matches large scale of extent with comprehensiveness of description (including *built-up density* and other 68 metrics of urban form) and detail of information (intra-urban, in fact down to the individual building level): this is, to our knowledge, unprecedented in this area of literature. We do so by utilising a novel toolkit (*momepy*) (Fleischmann, 2019), which allows both large-scale and rich "urban morphometrics" analysis starting from an extremely parsimonious set of input information: buildings and street network. Urban form is then related to COVID-19 cases and deaths up to May 2020 at the neighbourhood level.

## Datasets

The quantitative description of London's urban form and COVID-19 deaths and cases are obtained from four data sources: Ordnance Survey (OS) MasterMap, OS Open Roads, the underlying data of the COVID-19 Deaths Mapping Tool provided by the Greater London

Authority (GLA) and the number of contagions from coronavirus.data.gov.uk, a UK public sector website maintained by the Government Digital Service that provides government data of public utility. These are presented in more detail next.

OS MaterMap[1] is a set of geodatasets produced and kept updated by Ordnance Survey, the official mapping agency of the UK. They contain vectorial representations of various real-world physical features of the UK, such as buildings, green areas and water bodies. Each feature is a polygon, representing the area on the ground that the feature covers, in the British National Grid (BNG) coordinate system. For this study, two layers are required: Topography, with information on building footprints, and Building Height Attribute, with information on building heights.

OS Open Roads[2] is a geodatabase produced and kept updated by OS, in which each feature is a link representing a road centreline. OS Open Roads includes roads classified by the National or Local Highway authority (e.g., A roads, B roads) and unclassified paths, making up UK's road network.

The underlying data of the COVID-19 Deaths Mapping Tool[3] contains records of COVID-19 death rates for 1k residents for officially designated census areas, i.e., Middle Layer Super Output Areas (MSOAs), from the beginning of the pandemic (March 2020) until May 2020. It also contains further socioeconomic, ethnic and health indicators: proportion of population (PoP) over age 70, PoP with risk jobs, PoP with insecure jobs, PoP which is BAME, PoP with Black ethnicity, PoP with Pakistani or Bangladeshi ethnicity, PoP with Indian ethnicity, proportion of under-16s living in disadvantaged households, PoP with hypertension, PoP with obesity, PoP with diabetes, PoP with asthma, PoP with coronary heart disease and the Index of Multiple Deprivation (IMD) 2019.

The COVID-19 cases dataset reports the number of new cases among UK residents (people who have had at least one positive COVID-19 test result) in each MSOA, within 7-day rolling periods, starting from March 1st, 2020, and until the present day (06 June 2022). To estimate the total number of cases in each MSOA, we sum the number of cases in all the 7-day windows up to May 2020, for comparability with the previous dataset. We normalized the number of cases by the estimated population of area residents in year 2020.

---

[1] https://www.ordnancesurvey.co.uk/business-government/products/mastermap-topography
[2] https://www.ordnancesurvey.co.uk/business-government/products/open-map-roads
[3] https://data.london.gov.uk/download/covid-19-deaths-mapping-tool/9e3a5224-dc33-468d-b723-782be2b00967/underlying_data_2020_06_01.xlsx

# Methodology

## Computation of morphometrics

London's urban form is comprehensively described through a set of 69 urban morphometric descriptors, computed through *momepy*, an open-source Python-based tool for morphological analysis (Fleischmann, 2019). *momepy* allows to measure three main morphological components, building, streets and plot (Moudon, 1997), via five main properties: dimension, distribution, shape, intensity and connectivity. These quantify aspects of the single morphological elements and their spatial relationships. *Dimension* is about measuring the main physical aspects. It is quantified, for example, through building footprint, building height and street length. *Distribution* looks at their positioning in space. For instance, it is measured through building's cardinal orientation, building's street alignment and street edge[4] permeability. *Shape* quantifies the degree of physical complexity, e.g. plot compactness and street linearity. *Intensity* quantifies density in many ways, including buildings per meter of street, plot coverage ratio and floor area ratio; the latter, in particular, is a conventionally accepted measure of *built-up density*. *Connectivity* describes the configuration of the street network, as measured by, for example, local meshedness (the extent to which a street layout resembles more a grid or a tree-like structure), local closeness (assessing simultaneously connectivity and density of street segments) or proportion of 4-way intersections.

Since the plot is indeed defined and interpretated differently in different contexts (Kropf, 1997), *momepy* generates a new plot-proxy entity from building footprints and streets named "morphological cell", via Voronoi tessellation-based partitioning of space (Fleischmann et al., 2020). Since the 69 morphometrics are computed for buildings, streets and plots and COVID-19 deaths and cases are given for MSOAs, to make the analysis possible, the former is aggregated at the level of the latter through mean statistic. The full list of morphometrics is provided in Table S1 in the Supplementary Material, while formulas can be found in (Fleischmann et al., n.d.).

## Building the linear models

The relationship between London's urban form and COVID-19 effects is explored by building a linear regression model using the control variables included in the COVID-19 Deaths Mapping Tool and distance to centre (i.e., City of London), controlling for locational attributes of neighbourhoods in relation to the metropolitan context, and then by building a second linear regression model using relevant morphometrics with the residuals of the first model as target variable.

---

[4] A street edge is the built front composed by all the plots (and buildings) facing a street (Venerandi et al., 2017).

**Normalisation and scaling.** Since the statistical analysis used in this study is based on linear regression, which requires input variables to be normally distributed, the 69 morphometrics, socioeconomic indexes, pre-existing health conditions and COVID-19 deaths and cases are transformed via the Yeo-Johnson power transformation (Yeo and Johnson, 2000) to have distribution functions as close as possible to the normal. Furthermore, since regression coefficients must be comparable for interpretability, the normalised morphometrics are further transformed in z-scores, i.e. the difference between raw score and population mean, divided by the population standard deviation. A 0 z-score means that the value is aligned with the population mean; positive z-scores represent standard deviations above these values, while negative ones correspond to standard deviations below the mean.

**Feature selection.** A common issue in linear regression is the use of cross-correlated explanatory variables. This usually leads to overfitting and regression coefficients with inflated values and unexpected signs. To counteract this issue, two techniques are used conjointly to select the most descriptive variables to be used in the control models and in those with features of urban form: recursive feature elimination (RFECV) (Guyon et al., 2002) and hierarchically clustered cross-correlation matrix. The former fits a model and removes the weakest predictors until an optimal number is reached, in a cross-validated environment. The latter consists of a square matrix showing all possible pairwise Pearson's correlations among variables, organised through a hierarchical clustering algorithm using correlation coefficients as distance measures. The two techniques are used as follows: if RFECV selects variables which performed poorly once inserted in the model (e.g. particularly small and/or not statistically significant standardised regression coefficients), the hierarchically clustered cross-correlation matrix is used to determine the optimal subset of variables, by selecting one for each of the main branches below the second bifurcation of the respective dendrogram. A further iteration of RFECV can then be used to narrow down the selection and obtain a parsimonious and highly descriptive set of variables to be used in the model.

**Linear regression.** The main concept of linear regression is to explain the variation in a given variable (dependent) via a linear function of a set of other variables (explanatory). In its most general terms, the linear regression can be expressed as:

$$Y_i = a + \sum_k X_{ik} \beta_k + e_i$$

Where $Y_i$ is the dependent variable and $X_{ik}$ is a set of covariates used to explain $Y_i$. $\beta_k$ are the regression coefficients, that represent which way (positively or negatively) and to what extent each variable is related to $Y_i$. $a$ is the constant term, corresponding to the average value of $Y_i$ when all other variables are zero. Finally, $e_i$ is the error term (also called residuals), capturing elements that influence $Y_i$ but are not included among the explanatory variables. One of the main advantages of linear regression is that it isolates the distinct effects that each of such variables has on the dependent one, while "controlling" for the other in the model.

When modelling spatial variables, residuals may be spatially autocorrelated. This, in turn, violates the independent and identically distributed assumption of linear regression. To address this, the Moran's test (Cliff and Ord, 1981), a null hypothesis testing technique for the detection of spatial autocorrelation, can be used to assess the residuals, having previously built a spatial weight matrix for the observations. If the residuals show statically significant traces of spatial autocorrelation, Lagrange multiplier (LM) tests for the error and lagged models (Anselin et al., 1996) can be performed to decide the best modelling solution. In most cases, one of the two tests is statistically non-significant thus the other solution must be retained. Both error and lagged model account for spatial effects, however, in different ways. In the former (Arraiz et al., 2010), such effects are accounted for by computing a spatially lagged version of the error term ($u_i$) from the spatial weight matrix ($w$). More specifically:

$$Y_i = a + \sum_k \beta_k X_{ki} + u_i$$

$$u_i = \lambda u_{lag-1} + e_i$$

where $u_{lag-1} = \sum_j w_{i,j} u_j$ and $\lambda$ is the coefficient expressing the average strength of spatial correlation among the errors. In the latter (Anselin, 1988), spatial effects are accounted for by introducing a spatial lag of the dependent variable ($\rho Y_{lag-i}$), computed via $w$, on the right side of the equation. More specifically:

$$Y_i = a + \rho Y_{lag-i} + \sum_k \beta_k X_{ki} + e_i$$

## Application

### Pre-processing

Target variables, i.e., COVID-19 deaths and cases as well as socioeconomic and health indicators are extracted from the COVID-19 Deaths and the gov.uk COVID-19 contagion datasets. The 69 morphometrics are first computed at the level of their original spatial units (i.e., building, street, morphological cell) and then aggregated for MSOAs by calculating their means. For matter of brevity, only the target variables and a selection of four morphometrics are presented in Figure 1. Finally, the Yeo-Johnson power transformation is applied to all variables to render distribution functions the closest possible to the normal and z-scores are computed for each of them.

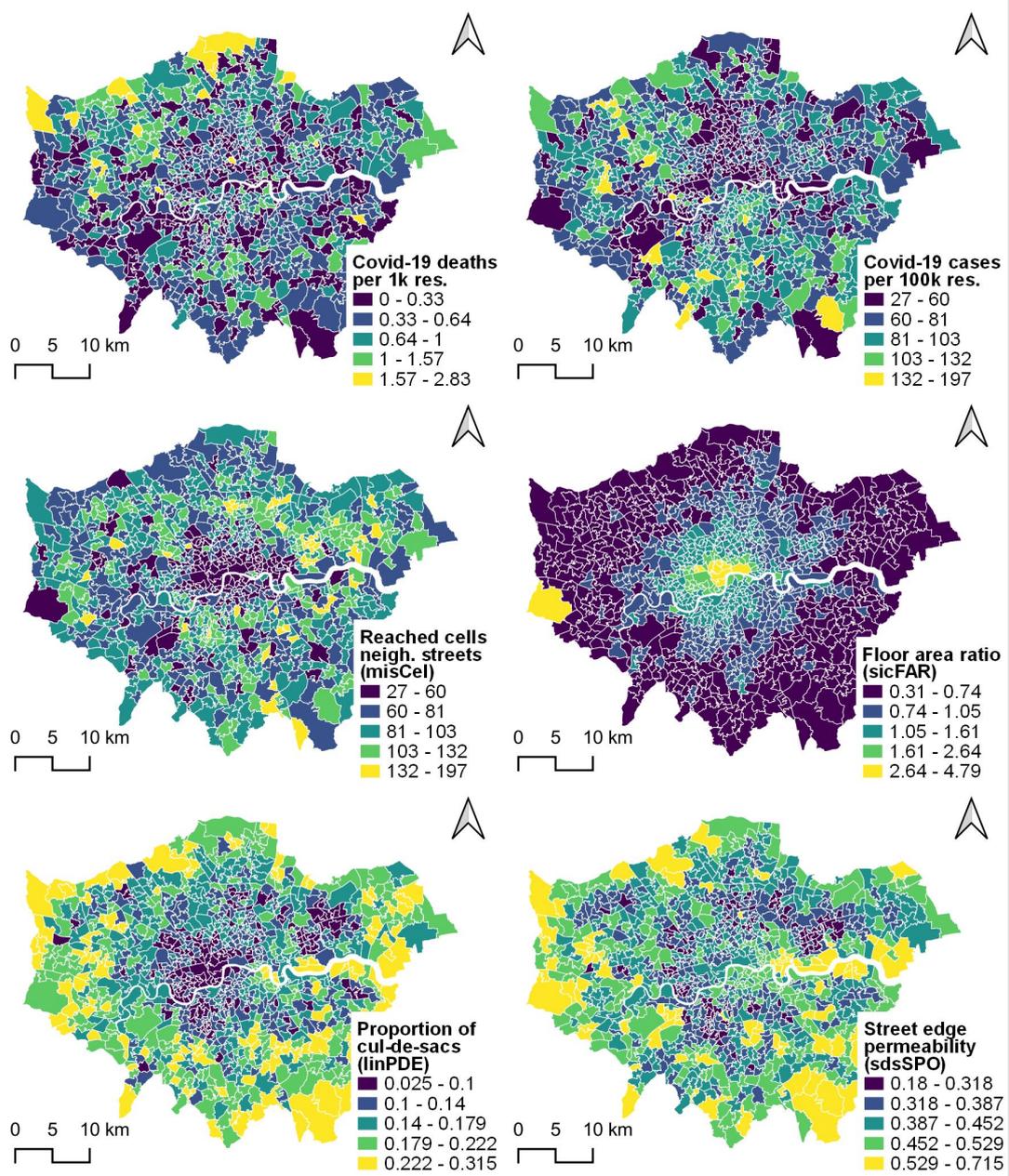

*Figure 1. COVID-19 deaths per 1k residents (first row, left), cases per 100k residents (first row, right) and examples of four morphometrics (second and third rows).*

## Control models for COVID-19 deaths and cases

A first iteration of RFECV, with COVID-19 deaths as target and all socioeconomic, health and distance to centre indicators as explanatory variables, was unsuccessful as the model performed poorly and several regression coefficients were not statically significant. Thus, a hierarchically clustered cross-correlation matrix was produced (Figure S1 in the Supplementary Material) from all possible pairwise combinations of control variables and a compact set was selected based on cluster membership and correlation levels, as explained in the Methodology. Due to the strong correlations with the other variables in the left branch, IMD score was selected as representative of the left branch. PoP with Indian ethnicity and PoP over age 70 were selected as main variables of the other two main branches as the former was the only representative of the middle branch, while the latter largely summarised the information of its cluster, including distance to centre, and also added the age factor. For the sake of completeness, regressions with distance to centre for both COVID-19 deaths and cases are nevertheless included in the Supplementary Material (Tables S2 and S3).

The same procedure was applied to identify an optimal subset of explanatory variables to model COVID-19 cases. In this case, RFECV was successful in identifying an optimal subset, i.e. PoP over age 70, PoP with Black ethnicity and PoP with diabetes. Both sets were regressed against their respective target variables (Table 1). A nearest neighbour weights matrix based on three nearest neighbours was introduced in both models to allow the computation of the Moran's test on the residuals.

For what concerns the model for COVID-19 deaths, the selected control variables explained 19% of the variance, with PoP over age 70 holding the strongest positive association (0.47), followed by the IMD score (0.42) and PoP with Indian ethnicity (0.21). Spatial autocorrelation in the residuals was non-statistically relevant (Moran's $I$ = 0.04, p-value = 0.07). Results point to the fact that more deaths tend to be associated with an older and more fragile population (with pre-existing health conditions as shown in Figure S1), lower socioeconomic statuses and greater proportions of people of Indian ethnicity. While the explanation for the former is more straightforward, interpretations for the latter two are lesser so. IMD seems to encapsulate large part of the remaining possible causes, i.e., overcrowding (one of its subindexes), pre-existing health condition (PoP with diabetes, obesity) and risk and insecure jobs (PoP with at risk jobs, insecure jobs), which are, in turn, associated with more face-to-face interactions and thus more chances of being infected and die of COVID-19. Furthermore, we hypothesise that London residents of Indian ethnicity are more affected due to pre-existing health conditions (see cross-correlations with PoP with diabetes, hypertension, and coronary heart disease in Figure S1). A further reason might be a lifestyle associated with more communal living, leading to more social interactions, e.g., large gatherings for family, cultural or religious purposes (Quadri, 2020; Yi et al., 2021), which, in turn, might be associated with a stronger virus propagation and consequently higher chances of death.

*Table 1. The control models for COVID-19 deaths per 1k residents (top) and cases per 100k residents (bottom).*

| Variable | Coefficient | Std. Error | t-Statistic | Probability |
|---|---|---|---|---|
| CONSTANT | 0.000 | 0.029 | 0.000 | 1.000 |
| Proportion of population (PoP) over age 70 | 0.467 | 0.036 | 12.833 | 0.000 |
| IMD score | 0.419 | 0.037 | 11.448 | 0.000 |
| Proportion of population (PoP) with Indian ethnicity | 0.214 | 0.029 | 7.361 | 0.000 |
| | | | R-squared: | 0.192 |
| | | | Adjusted R-squared: | 0.190 |
| | | | Prob(F-statistic): | 4.278e-45 |
| | | | Moran's I: | 0.041 |
| | | | p-value: | 0.071 |

| Variable | Coefficient | Std. Error | z-Statistic | Probability |
|---|---|---|---|---|
| CONSTANT | 0.002 | 0.039 | 0.058 | 0.954 |
| Proportion of population (PoP) over age 70 | 0.421 | 0.034 | 10.646 | 0.000 |
| Proportion of population (PoP) with Black ethnicity | 0.404 | 0.042 | 6.161 | 0.000 |
| Proportion of population (PoP) with diabetes | 0.227 | 0.037 | 7.643 | 0.000 |
| lambda | 0.324 | 0.030 | 10.737 | 0.000 |
| | | | Pseudo R-squared: | 0.226 |

For what concerns the model for COVID-19 cases, linear regression showed spatial autocorrelation in the residuals (Moran's *I* = 0.22, p-value = 0.0000) and the LM tests pointed to the use of the spatial error model (p-values of robust LM lag and error tests were 0.26 and 0.0001, respectively). Having implemented such a model, the selected control variables together with the parameter of the error term (lambda) reached a pseudo-R squared of 0.23, with PoP over age 70 holding the strongest positive association (0.42), followed by PoP with Black ethnicity (0.40), lambda (0.32) and PoP with diabetes (0.23). The result confirms, to a certain extent, the outcome of the first model, reiterating the fact that age and ethnicity – though Black in this case – are important predictors of the target variable. The other relevant explanatory factor is PoP with diabetes which is a known risk factor (Jeong et al., 2020).

## Models with selected morphometrics

A first iteration of RFECV, with the residuals of the first control model as target and the 69 morphometrics as explanatory variables, was unsuccessful as the model with the identified set proved to perform very poorly and featured regression coefficients not statistically significant. Thus, five hierarchically clustered cross-correlation matrices were computed for each of the five categories of morphometrics, i.e., dimension, distribution, shape, intensity and connectivity (Figures from S2 to S6 in the Supplementary Material), to select the most relevant among them. One representative was chosen for each of the main branches, obtained from cutting the dendrogram below the second bifurcation. The final set of metrics is as follows. Dimension: building's volume (sdbVol); number of morphological cells reached by neighbouring streets (misCel); street canyon width (sdsSPW). Distribution: street edge permeability (sdsSPO); block's cardinal orientation (lteOri); block's weighted neighbours (lteWNB). Shape: building's elongation (ssbElo); building's volumetric compactness (ssbVFR); height to width ratio (sdsSPR). Intensity: floor area ratio or *built-up density* (sicFAR); buildings per meter of street (sisBpM); density of morphological cells at block level (lieWCe). Connectivity: proportion of cul-de-sacs (linPDE); local closeness (lcnClo); proportion of 3-way intersections (linP3W). To further reduce the number of morphometrics, RFECV was used again with the residuals of the first control model as target variable. Six highly representative morphometrics were identified: linPDE, lcnClo, sdsSPO, lteWNB, sicFAR and sdsSPR. The same process was carried out for selecting morphometrics to regress against the residuals of the model for COVID-19 cases. A first iteration of RFECV did not provide satisfactory results. However, a second iteration with the morphometrics selected via the correlation matrices output a relevant set: sdbVol, lteOri, sicFAR, ssbElo and sdsSPR. These two sets were then regressed against the residuals of the two control models. The results of both models are presented in Table 2.

Outcomes show that the selected morphometrics have a marginal role in explaining the variance left unexplained by the control models, i.e., 4%. Spatial autocorrelation is non-statistically relevant in the first (Moran's $I$ = 0.01, p-value = 0.5946), while it is in the second (Moran's $I$ = 0.20, p-value = 0.0000). The spatial lag model is thus implemented due to the statistical relevance of the robust LM lag test (0.0014) and non-relevance of the robust LM error test (0.0524). In terms of regression coefficients, urban density measured as floor area ratio (sicFAR) is present and statistically significant in both models. It shows the strongest regression coefficients among the selected morphometrics (-0.21 and -0.28): it is negative, meaning that less *built-up density* is associated with more COVID-19 deaths and cases.

*Table 2. Models with selected morphometrics for COVID-19 deaths (top) and cases (bottom).*

| Variable | Coefficient | Std.Error | t-Statistic | Probability |
|---|---|---|---|---|
| CONSTANT | 0.000 | 0.028 | 0.000 | 1.000 |
| Proportion of cul-de-sacs (linPDE) | 0.170 | 0.051 | 3.326 | 0.001 |
| Local closeness (lcnClo) | -0.132 | 0.031 | -4.289 | 0.000 |
| Street profile openness (sdsSPO) | -0.181 | 0.038 | -4.752 | 0.000 |
| Blocks' granularity (lteWNB) | 0.095 | 0.056 | 1.694 | 0.091 |
| Floor area ratio (sicFAR) | -0.212 | 0.089 | -2.364 | 0.018 |
| Height to width ratio (sdsSPR) | 0.182 | 0.084 | 2.161 | 0.031 |
| | | | R-squared: | 0.042 |
| | | | Adjusted R-squared: | 0.037 |
| | | | Prob(F-statistic): | 1.599e-07 |
| | | | Moran's I: | 0.009 |
| | | | Prob(F-statistic): | 0.595 |

| Variable | Coefficient | Std.Error | z-Statistic | Probability |
|---|---|---|---|---|
| CONSTANT | 0.003 | 0.027 | 0.116 | 0.908 |
| Building volume (sdbVol) | 0.128 | 0.037 | 3.465 | 0.001 |
| Blocks' cardinal orientation (lteOri) | 0.027 | 0.034 | 0.784 | 0.433 |
| Floor area ratio (sicFAR) | -0.279 | 0.098 | -2.847 | 0.004 |
| Building elongation (ssbElo) | -0.081 | 0.043 | -1.890 | 0.059 |
| Height to width ratio (sdsSPR) | 0.144 | 0.081 | 1.778 | 0.075 |
| W_residuals | 0.702 | 0.211 | 3.327 | 0.001 |
| | | | Pseudo R-squared: | 0.127 |
| | | | Spatial Pseudo R-squared: | 0.036 |

In the model for COVID-19 deaths, linPDE and lcnClo showed positive and negative regression coefficients respectively (0.17 and -0.13), meaning that less street network connectivity and density is related to more deaths. sdsSPO was negatively associated (-0.18), meaning that less permeable street edges are related to more deaths. sdsSPR showed a positive $\beta$ (0.18), meaning that greater ratios (corresponding to smaller street sections) are positively associated with the target variable. lteWNB is not reported as not statistically significant (p-value > 0.05). In the model for COVID-19 cases, sdbVol was positively associated with more cases (0.13). However, the combination of this metric with sicFAR, showing a negative $\beta$, points to a density not distributed in space, but rather concentrated, typical of sprawling and modernist layouts, i.e., free-standing buildings on large plots. ssbElo, lteOri and sdsSPR are not reported as not statistically significant.

The picture emerging from the interpretation of the regression coefficients of both models is that of a COVID-19 exposed London area with low levels of street network connectivity and *built-up density*, featuring a porous urban fabric mainly characterized by semi-detached, small terraced houses, small street sections and not too permeable edges, dotted by larger buildings, such as single houses, pavilions and tower blocks sitting free-standing on cells of significant size: overall, a typical urban fringe type of development. To validate this outcome and offer a visual representation, we first selected the top MSOAs for COVID-19 deaths and cases via natural breaks binning; second, among these, we picked the top two MSOAs for which our models best performed (i.e., where the errors were the smallest) (Figure 2). Sample street-view images for the same MSOAs are presented in Figure S7 in the Supplementary Material.

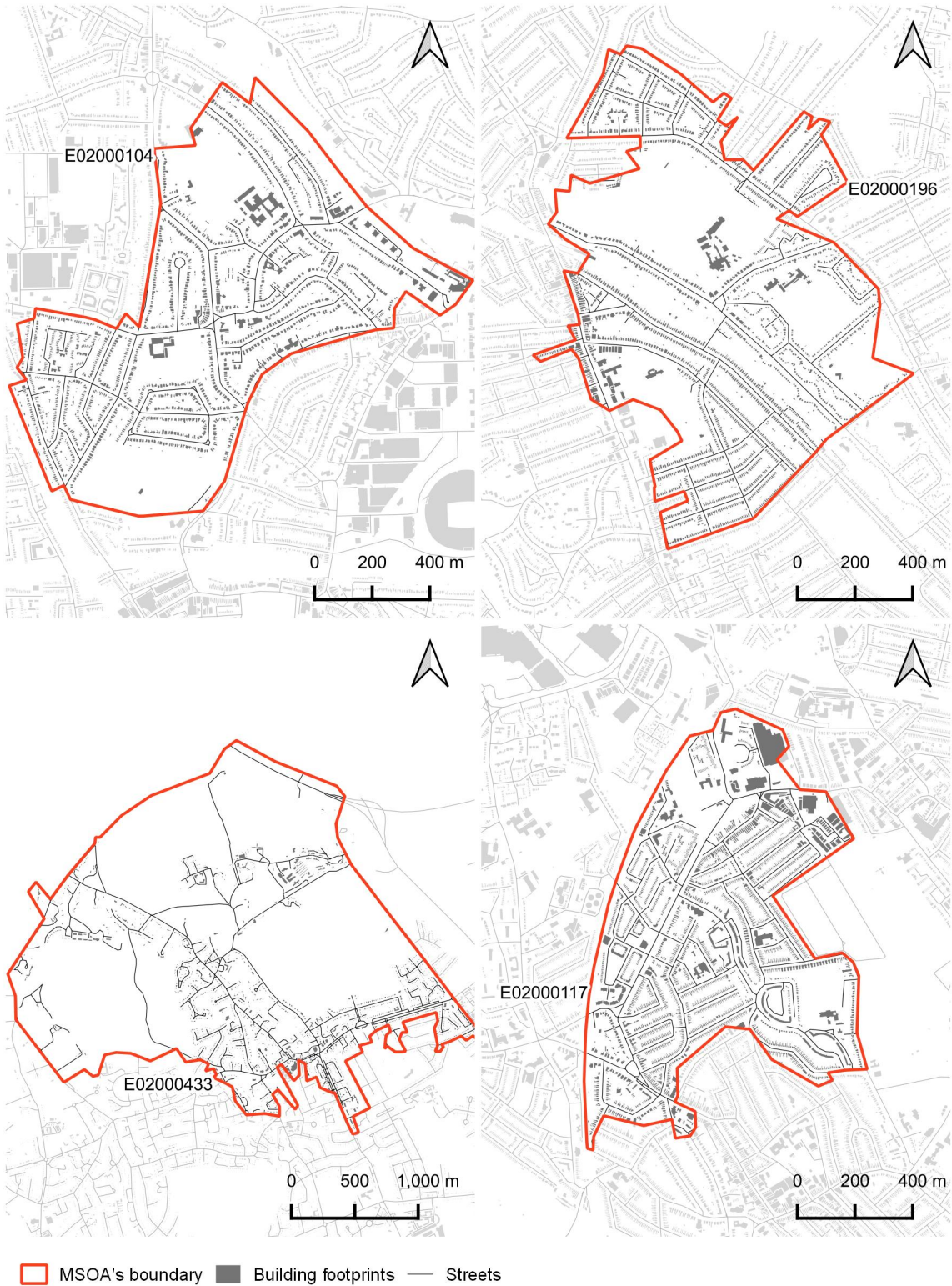

*Figure 2. Worst affected MSOAs in terms of COVID-19 deaths (top row) and cases (bottom row), where models perform the best. Buildings: OS MasterMap. Streets: OS Open Roads.*

## Discussion

Compared to previous studies focusing on the relationship between spread of COVID-19 and population density (Carozzi, 2020; Fang and Wahba, 2020; Hamidi et al., 2020), this work affords a finer level of spatial granularity (i.e., neighbourhoods rather than whole cities) and investigates density through the lens of urban morphology. Furthermore, in line with the studies mentioned above, it suggests that, in the relatively well controlled environment of Greater London, where the issue of overcrowding is not of particular concern (i.e., only 7.5% of households are classified overcrowded),[5] the negative impacts of the COVID-19 pandemic have more to do with socioeconomics, pre-existing health conditions, and age than other factors. The marginal contribution of the configuration of the built environment and, in particular, of built-up density, which is negatively associated with both COVID-19 deaths and cases, contradicts popular unproven assumptions that "density" is detrimental for communicable respiratory diseases, due to people living in tight urban environments. Future work may consider replicating the very same methodology proposed in this paper to case studies where overcrowding is much more pronounced, to quantify the relative weights of this phenomenon and relevant descriptors of urban form in relation to the negative impacts of COVID-19.

This study is affected by four main limitations. First, we focused our analysis on the first wave of the COVID-19 pandemic because area-level data on COVID deaths is limited to that period. As the pandemic evolved, changes in the public response and the rollout of the vaccination campaign might have affected the relative weight of correlates with COVID-19 cases and deaths. Second, the recorded COVID-19 prevalence across areas depends on the number of tests performed. However, no data on number of tests is available at MSOAs level. Studies conducted in the US during the first COVID-19 wave showed that access to COVID-19 testing was geographically uneven at country level due to multiple factors including diagnostic testing capacity (CDC COVID-19 Response Team, 2020), but those differences did not hold at city-level (Schmitt-Grohé et al., 2020). A third limitation regards generalizability. While the results hold for Greater London, they might not be valid in other contexts. However, the replicability of the methodology opens to future work in this direction. Fourth, results of regression analyses do not imply causation.

## Conclusion

With the COVID-19 pandemic, the equation "less density equals healthier cities" is back in the centre-stage of the debate in urban design, even though empirical proofs of this lack. Existing studies on the relationship between density and COVID-19 are largely inconclusive, mainly focus on entire metropolitan regions and do not consider aspects of urban form. To better ascertain the role of the latter (including built-up density) in relation to COVID-19, we analysed the relationship between a comprehensive set of descriptors of urban form and both COVID-19 deaths and cases, in Greater London, while controlling for socioeconomics, co-morbidity and age. Results show that the largest part of the variance is explained by the control factors, while descriptors of urban form, including built-up density, play a very marginal role. Moreover, insofar as urban form is concerned, the latter is indeed the most relevant factor, but is *inversely* related to it. Visual inspection of MSOAs where models perform the best reveals fabrics typical of urban fringes, characterised by low-density housing dotted by free-standing larger buildings (e.g., pavilions, tower blocks), and poor street network connectivity.

---

[5] https://data.london.gov.uk/dataset/overcrowded-households-borough?q=overcrowding

# Supplementary Material for "Urban form and COVID-19 cases and deaths in Greater London: an urban morphometric approach"

Alessandro Venerandi*, Luca Maria Aiello[†], Sergio Porta*

* Urban Design Studies Unit (UDSU), Department of Architecture, University of Strathclyde, Glasgow, UK. 75 Montrose St, Glasgow G1 1XJ, UK
[†] Department of Computer Science, IT University of Copenhagen, Copenhagen, DK Rued Langgaards Vej 7, DK-2300 Copenhagen S, Denmark

Corresponding author's email: alessandro.venerandi@strath.ac.uk

Table S1. List of the 69 morphometrics, alongside spatial element, scale, spatial context, and conceptual category.

| character | element | scale | context | category |
|---|---|---|---|---|
| area | building | S | building | dimension |
| height | building | S | building | dimension |
| volume | building | S | building | dimension |
| perimeter | building | S | building | dimension |
| courtyard area | building | S | building | dimension |
| form factor | building | S | building | shape |
| volume to façade ratio | building | S | building | shape |
| circular compactness | building | S | building | shape |
| corners | building | S | building | shape |
| squareness | building | S | building | shape |
| equivalent rectangular index | building | S | building | shape |
| elongation | building | S | building | shape |
| centroid - corner distance deviation | building | S | building | shape |
| centroid - corner mean distance | building | S | building | shape |
| cardinal orientation | building | S | building | distribution |
| street alignment | building | S | building | distribution |
| cell alignment | building | S | building | distribution |
| longest axis length | tessellation cell | S | tessellation cell | dimension |
| area | tessellation cell | S | tessellation cell | dimension |
| circular compactness | tessellation cell | S | tessellation cell | shape |
| equivalent rectangular index | tessellation cell | S | tessellation cell | shape |
| cardinal orientation | tessellation cell | S | tessellation cell | distribution |
| *coverage area ratio* | tessellation cell | S | tessellation cell | intensity |
| *floor area ratio* | tessellation cell | S | tessellation cell | intensity |
| *length* | street segment | S | street segment | dimension |
| *width* | street profile | S | street segment | dimension |
| *height* | street profile | S | street segment | dimension |
| *height to width ratio* | street profile | S | street segment | shape |
| *openness* | street profile | S | street segment | distribution |
| *width deviation* | street profile | S | street segment | diversity |
| *height deviation* | street profile | S | street segment | diversity |
| *linearity* | street segment | S | street segment | shape |
| *area covered* | street segment | S | street segment | dimension |
| *buildings per meter* | street segment | S | street segment | intensity |
| *area covered* | street node | S | street node | dimension |
| *shared walls ratio* | adjacent buildings | M | adjacent buildings | distribution |
| *alignment* | neighbouring buildings | M | neighbouring cells (queen) | distribution |
| *mean distance* | neighbouring buildings | M | neighbouring cells (queen) | distribution |
| *weighted neighbours* | tessellation cell | M | neighbouring cells (queen) | distribution |
| *area covered* | neighbouring cells | M | neighbouring cells (queen) | dimension |
| *reached cells* | neighbouring segments | M | neighbouring segments | intensity |
| *reached area* | neighbouring segments | M | neighbouring segments | dimension |
| *degree* | street node | M | neighbouring nodes | distribution |
| *mean distance to neighbouring nodes* | street node | M | neighbouring nodes | dimension |
| *perimeter wall length* | adjacent buildings | L | joined buildings | dimension |
| *mean inter-building distance* | neighbouring buildings | L | cell queen neighbours 3 | distribution |
| *weighted reached blocks* | neighbouring tessellation cells | L | cell queen neighbours 3 | intensity |
| *area* | block | L | block | dimension |
| *perimeter* | block | L | block | dimension |
| *circular compactness* | block | L | block | shape |
| *equivalent rectangular index* | block | L | block | shape |
| *compactness-weighted axis* | block | L | block | shape |
| *cardinal orientation* | block | L | block | distribution |
| *weighted neighbours* | block | L | block | distribution |
| *weighted cells* | block | L | block | intensity |
| *local meshedness* | street network | L | nodes 5 steps | connectivity |
| *mean segment length* | street network | L | segment 3 steps | dimension |
| *cul-de-sac length* | street network | L | nodes 3 steps | dimension |
| *area covered* | street network | L | nodes 3 steps | dimension |
| *reached cells* | street network | L | segment 3 steps | dimension |
| *reached cells* | street network | L | nodes 3 steps | dimension |
| *reached area* | street network | L | nodes 3 steps | dimension |
| *node density* | street network | L | nodes 5 steps | intensity |
| *weighted node density* | street network | L | nodes 5 steps | intensity |
| *proportion of cul-de-sacs* | street network | L | nodes 5 steps | connectivity |
| *proportion of 3-way intersections* | street network | L | nodes 5 steps | connectivity |
| *proportion of 4-way intersections* | street network | L | nodes 5 steps | connectivity |
| *local closeness centrality* | street network | L | nodes 5 steps | connectivity |
| *square clustering* | street network | L | nodes within network | connectivity |

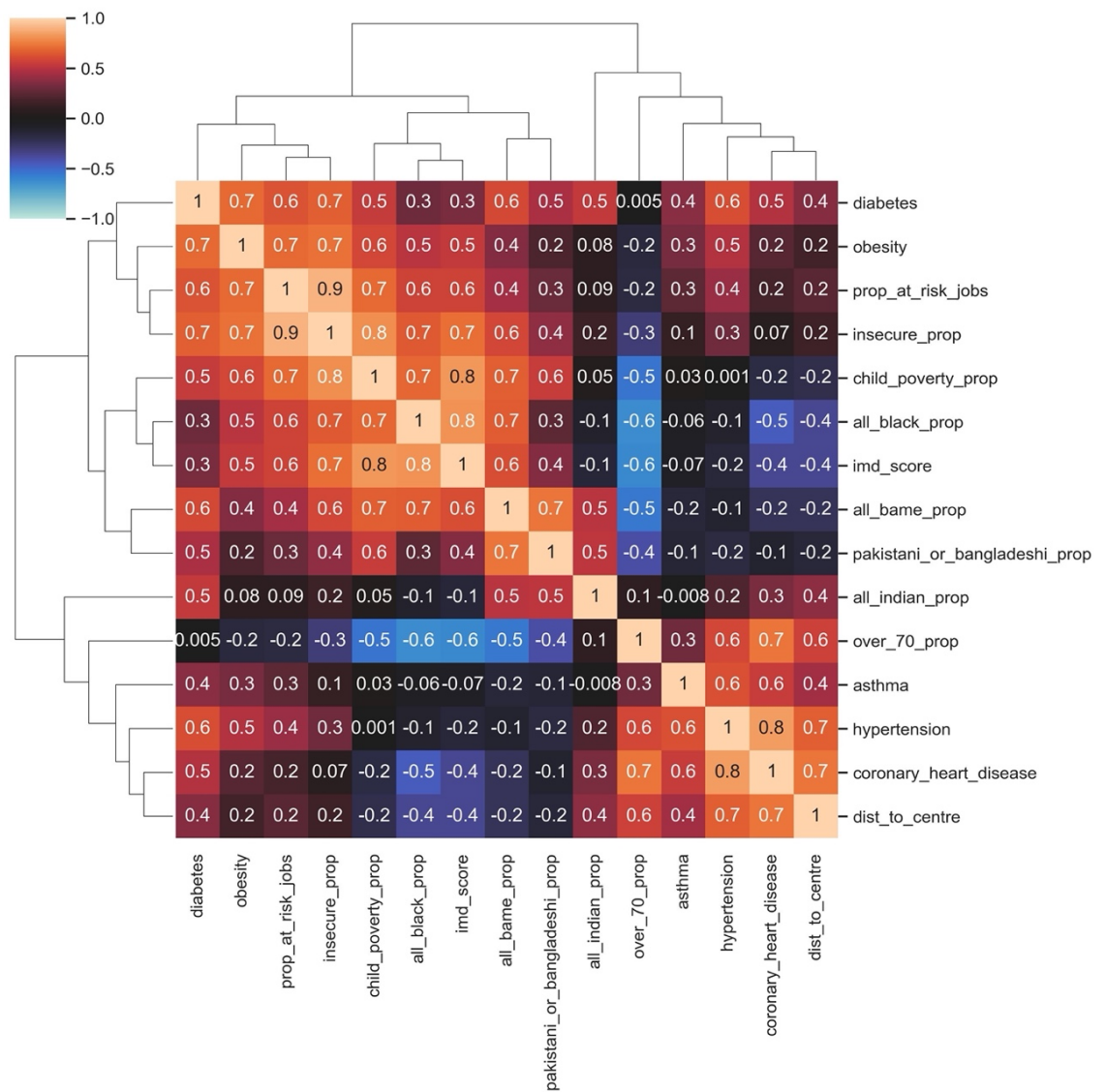

*Figure S1. Hierarchically clustered cross-correlation matrix of the control variables. Lighter warm hues correspond to stronger positive correlations. Black corresponds to no correlation. Lighter cold hues correspond to stronger negative correlations.*

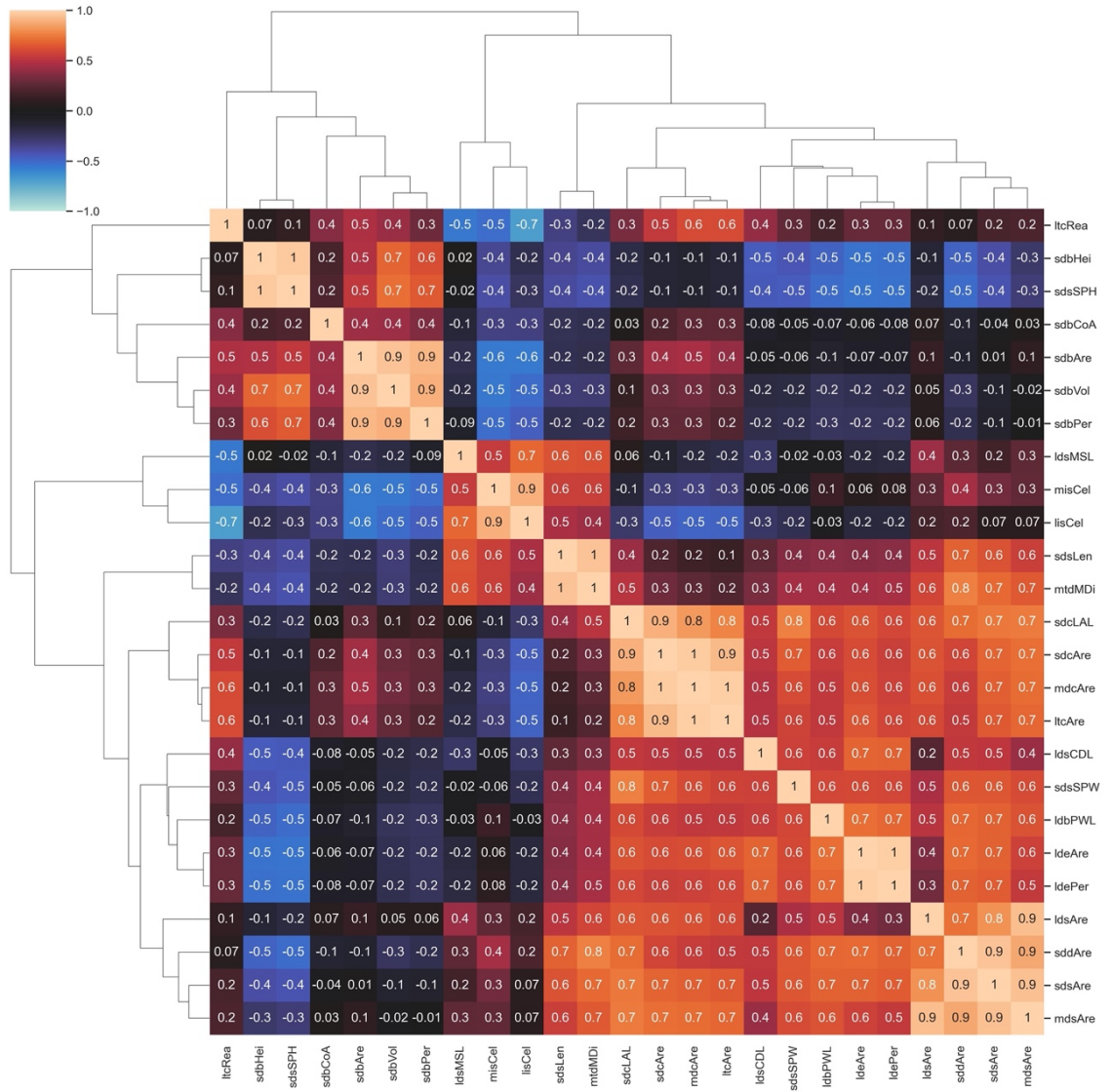

*Figure S2. Hierarchically clustered cross-correlation matrix of the morphometrics related to dimension. Lighter warm hues correspond to stronger positive correlations. Black corresponds to no correlation. Lighter cold hues correspond to stronger negative correlations.*

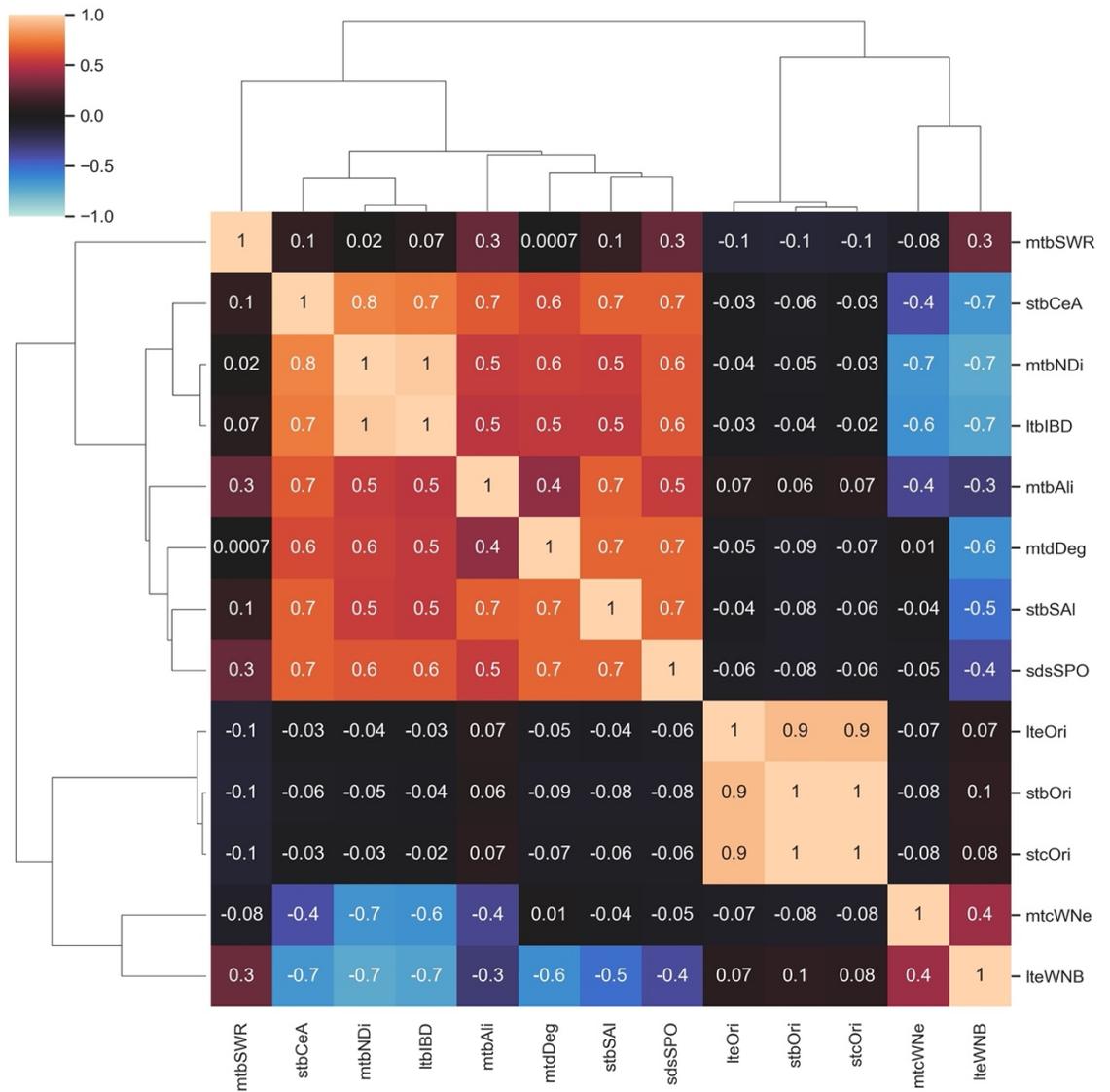

Figure S3. Hierarchically clustered cross-correlation matrix of the morphometrics related to distribution. Lighter warm hues correspond to stronger positive correlations. Black corresponds to no correlation. Lighter cold hues correspond to stronger negative correlations.

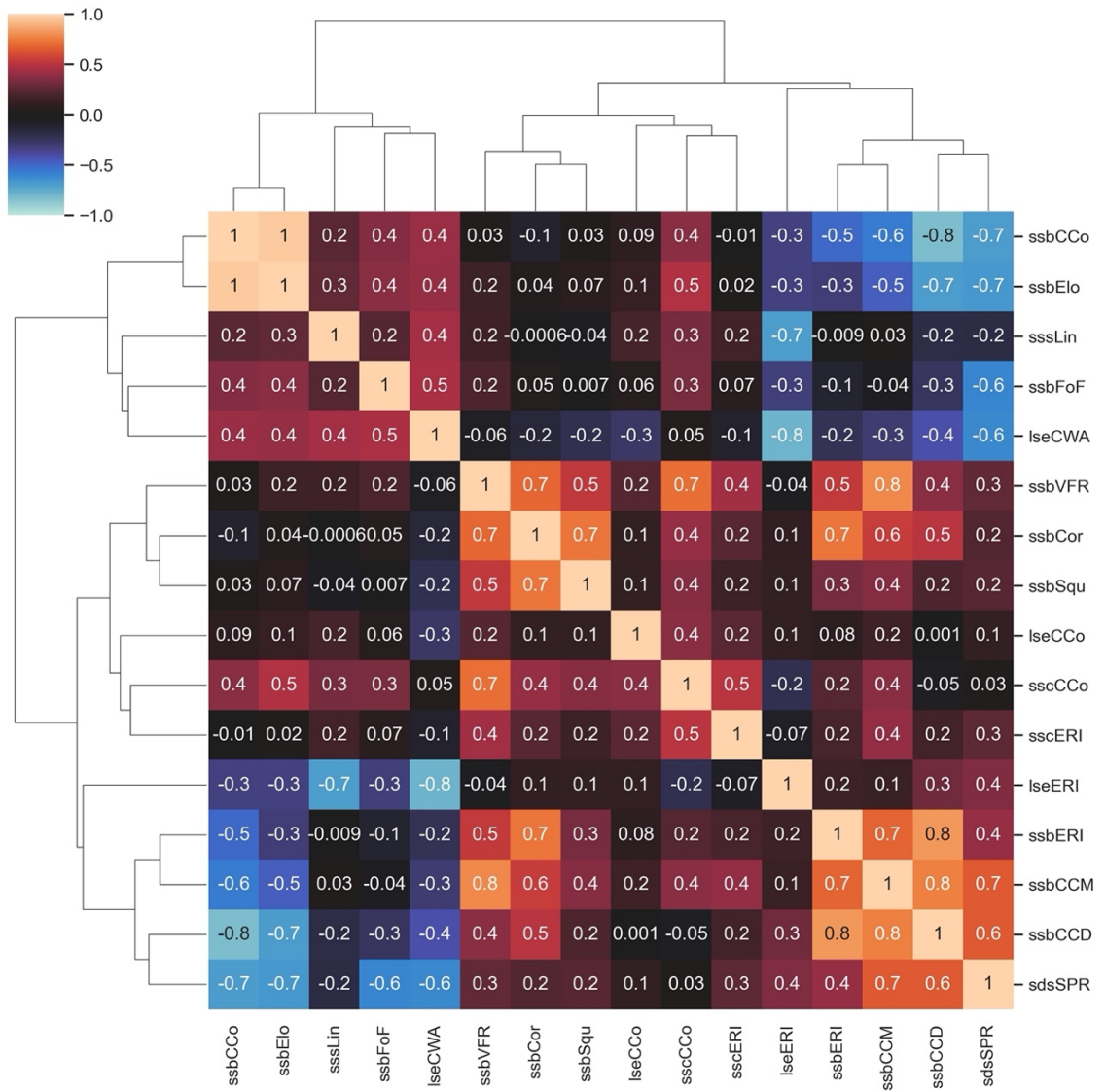

Figure S4. Hierarchically clustered cross-correlation matrix of the morphometrics related to shape. Lighter warm hues correspond to stronger positive correlations. Black corresponds to no correlation. Lighter cold hues correspond to stronger negative correlations.

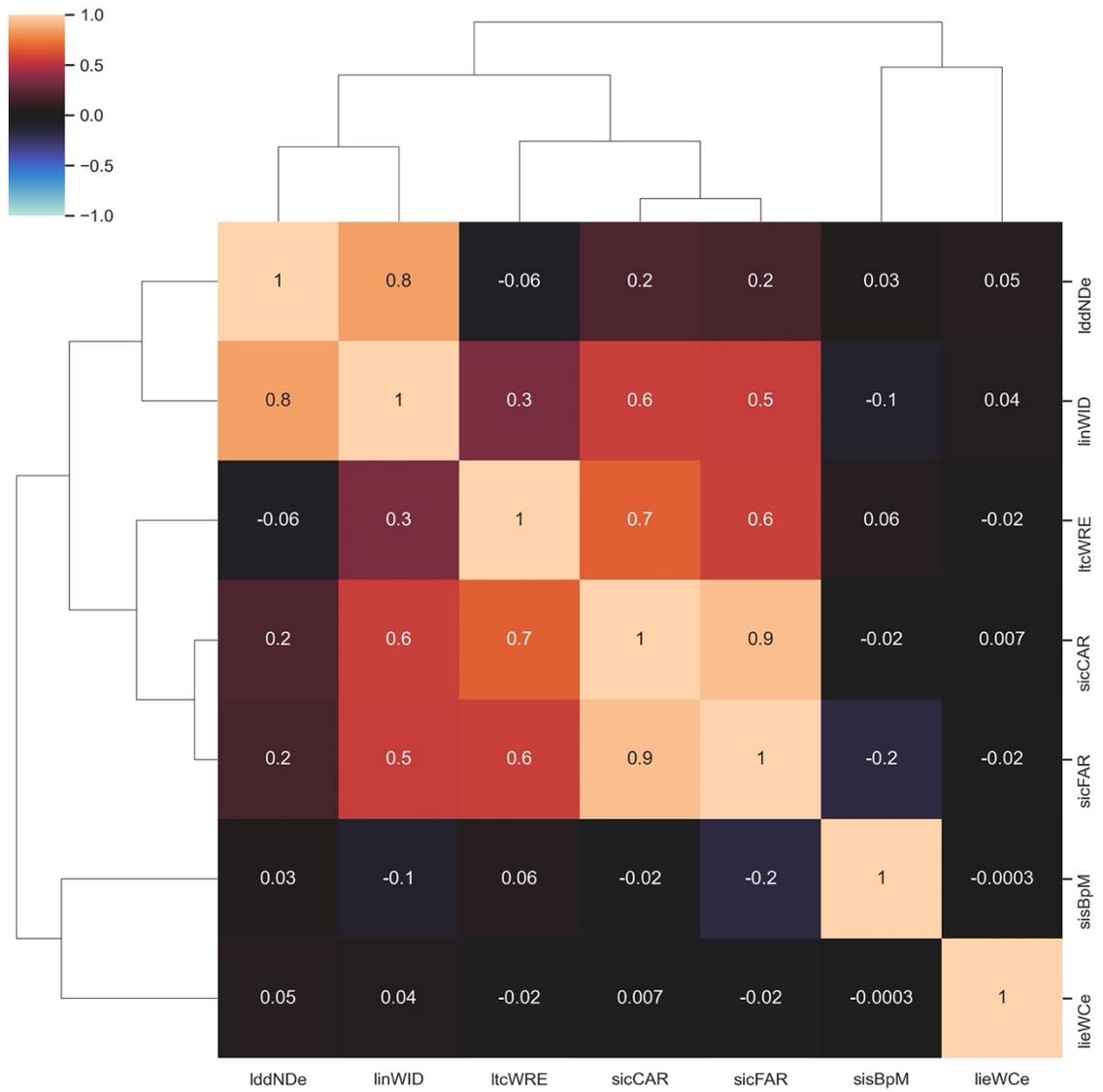

Figure S5. Hierarchically clustered cross-correlation matrix of the morphometrics related to intensity. Lighter warm hues correspond to stronger positive correlations. Black corresponds to no correlation. Lighter cold hues correspond to stronger negative correlations.

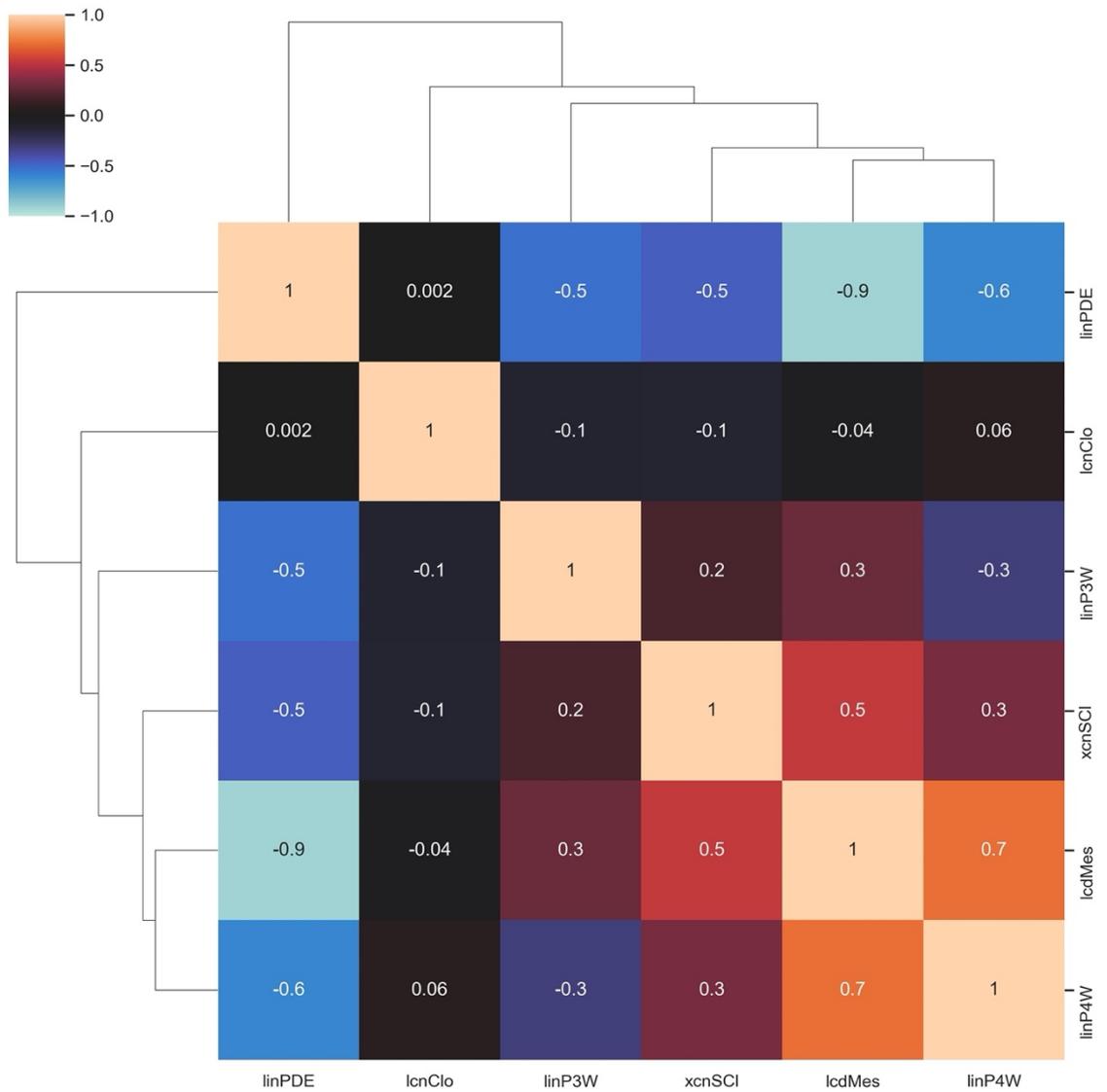

Figure S6. Hierarchically clustered cross-correlation matrix of the morphometrics related to connectivity. Lighter warm hues correspond to stronger positive correlations. Black corresponds to no correlation. Lighter cold hues correspond to stronger negative correlations.

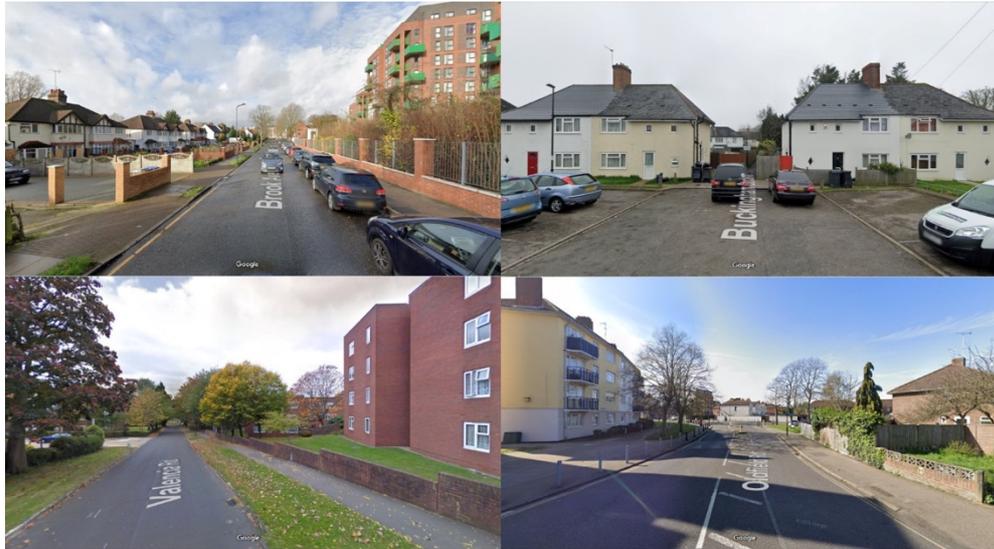

*Figure S7. Street views of worst affected MSOAs in terms of COVID-19-related deaths (top row) and cases (bottom row), where models perform the best. Source: Google Street View*

*Table S2. Regression model for COVID-19 deaths per 1k residents*

| Variable | Coefficient | Std.Error | t-Statistic | Probability |
|---|---|---|---|---|
| CONSTANT | 0.000 | 0.028 | 0.000 | 1.000 |
| Proportion of population (PoP) over age 70 | 0.533 | 0.043 | 12.46 | 0.000 |
| IMD score | 0.449 | 0.038 | 11.95 | 0.000 |
| Proportion of population (PoP) with Indian ethnicity | 0.216 | 0.033 | 6.527 | 0.000 |
| Distance to centre | -0.076 | 0.069 | -1.096 | 0.273 |
| Proportion of cul-de-sacs (linPDE) | 0.187 | 0.052 | 3.567 | 0.000 |
| Local closeness (lcnClo) | -0.120 | 0.037 | -3.255 | 0.001 |
| Street profile openness (sdsSPO) | -0.191 | 0.040 | -4.832 | 0.000 |
| Blocks' granularity (lteWNB) | 0.093 | 0.058 | 1.594 | 0.111 |
| Floor area ratio (sicFAR) | -0.197 | 0.092 | -2.144 | 0.032 |
| Height to width ratio (sdsSPR) | 0.141 | 0.092 | 1.525 | 0.128 |
| | | | R-squared: | 0.229 |
| | | | Adjusted R-squared: | 0.222 |
| | | | Prob(F-statistic): | 6.724e-49 |
| | | | Moran's I: | 0.009 |
| | | | p-value: | 0.528 |

*Table S3. Regression model for COVID-19 cases per 100k residents*

| Variable | Coefficient | Std.Error | z-Statistic | Probability |
|---|---|---|---|---|
| CONSTANT | -0.001 | 0.037 | -0.015 | 0.987 |
| Proportion of population (PoP) over age 70 | 0.410 | 0.049 | 8.386 | 0.000 |
| Proportion of population (PoP) with Black ethnicity | 0.432 | 0.045 | 9.652 | 0.000 |
| Proportion of population (PoP) with diabetes | 0.238 | 0.043 | 5.596 | 0.000 |
| Distance to centre | 0.198 | 0.069 | 2.856 | 0.004 |
| Building volume (sdbVol) | | | | |
| Blocks' cardinal orientation (lteOri) | 0.161 | 0.042 | 3.833 | 0.000 |
| Floor area ratio (sicFAR) | -0.280 | 0.101 | -2.765 | 0.006 |
| Building elongation (ssbElo) | -0.067 | 0.051 | -1.311 | 0.190 |
| Height to width ratio (sdsSPR) | 0.276 | 0.083 | 3.347 | 0.001 |
| W_residuals | 0.298 | 0.031 | 9.680 | 0.000 |
| | | | Pseudo R-squared: | 0.263 |